\documentclass{INTERSPEECH2023}

\usepackage{graphicx}
\usepackage{subcaption}
\usepackage{pdfpages}
\usepackage{tabularray}
\usepackage[textsize=tiny]{todonotes}

\usepackage{cite}
\interspeechcameraready

\title{On-Device Speaker Anonymization of Acoustic Embeddings for ASR based on Flexible Location Gradient Reversal Layer}

\name{Md Asif Jalal$^1$, Pablo Peso Parada$^1$, Jisi Zhang$^1$, Karthikeyan Saravanan$^1$, Mete Ozay$^1$, Myoungji Han$^2$, Jung In Lee$^2$, Seokyeong Jung$^2$}

\address{
  $^1$Samsung Research, UK\\
  $^2$AI R\&D Group, Samsung Electronics, Suwon, South Korea}
\email{mdasif.jalal@samsung.com}

\begin{document}

\maketitle
 
\begin{abstract}

Smart devices serviced by large-scale AI models necessitates user data transfer to the cloud for inference.
For speech applications, this  means transferring  private  user information, e.g., speaker identity.
Our paper proposes a privacy-enhancing framework that targets speaker identity anonymization while preserving speech recognition accuracy for our downstream task~-~Automatic Speech Recognition (ASR). 
The proposed framework attaches flexible \textit{gradient reversal based speaker adversarial layers} to target layers within an ASR model, where speaker adversarial training anonymizes acoustic embeddings generated by the targeted layers to remove speaker identity. 
We propose on-device deployment by execution of initial layers of the ASR model,  and transmitting anonymized embeddings to the cloud, where the rest of the model is executed while preserving privacy.
Experimental results show that our method efficiently reduces speaker recognition relative accuracy by 33\%, and improves ASR performance by achieving 6.2\% relative Word Error Rate (WER) reduction.

\end{abstract}

\vspace*{-5pt}\section{Introduction}
\label{sec::intro}

The increasing prevalence of voice driven human-computer interaction services in  appliances has raised concern with regard to voice privacy and personal information protection. These `smart' devices, ranging from cars to small watches, collect speech utterances and acoustic events for various downstream tasks or for training and evaluation in distributed settings \cite{GUPTA20181}. Speech utterances hold user information such as speaker identity, gender etc. Privacy preservation is of critical importance to protect reliability in private data sharing. 

Various privacy preservation methods for speech have been proposed in the literature. One solution is to manipulate speaker identity related features through feature perturbation~\cite{9102875}, voice normalisation~\cite{Qian2018HidebehindEV,9053868}, utterance slicing techniques~\cite{Maouche2022EnhancingSP}, and differential pitch anonymization~\cite{Shamsabadi2022DifferentiallyPS}. State-of-the-art methods employ neural based speech synthesizer or voice converter to generate speech where the speaker identity information has been removed~\cite{meyer2022speaker,meyer2023anonymizing}. However, these methods require employment of additional synthesis modules and are computationally expensive, which is unrealistic for on-device scenarios.

\begin{figure*}
    \centering
    \includegraphics[width=0.9\linewidth,height=0.3\linewidth, page=1,trim={0cm 7cm 0cm 0cm},clip]{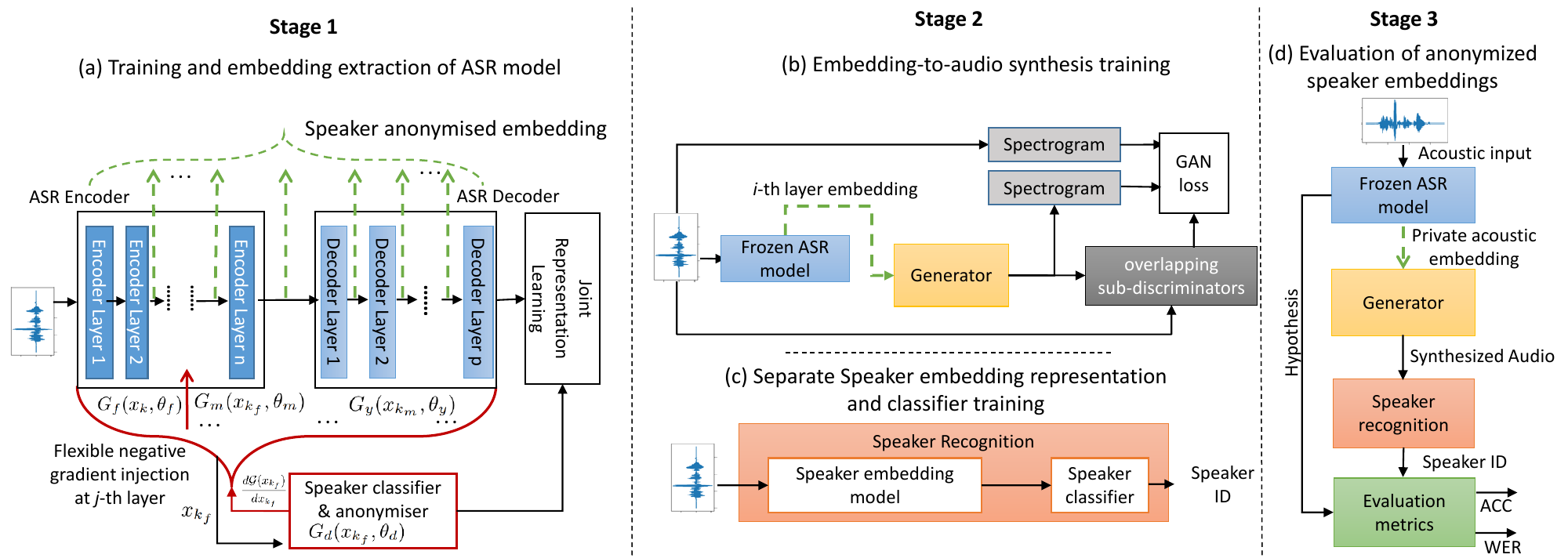}
    \vspace{-0.1cm}
    \caption{Our framework proposed for speaker anonymization and evaluation with acoustic embeddings.}
    \label{fig:arc}\vspace*{-20pt}
\end{figure*}

An alternative approach for speaker anonymization is to learn speech representations invariant to speaker conditions. Domain adversarial training trains a model to learn domain agnostic representations \cite{ganin2016domain}. Speaker based domain adversarial training has been effective for anonymizing latent representations of ASR models~(i.e., acoustic embeddings)~\cite{Srivastava2019PrivacyPreservingAR, Jaiswal2019PrivacyEM}. However, it was observed that speaker invariant representations resulted in a reduction of ASR performance \cite{Srivastava2019PrivacyPreservingAR}. 
Orthogonal to this, recent work by~\cite{zhou2022enhancing} discuss a method where adding speaker-labels and adaptive gradient scaling to domain adversarial training improves ASR performance. However, they do not target or discuss privacy.

In this paper, we propose a flexible gradient reversal based speaker anonymization framework, which learns speaker anonymous acoustic embeddings within an ASR model while preserving its accuracy/performance (as depicted in Stage 1 in Figure~\ref{fig:arc}). The initial layers of ASR models learn generic acoustic and prosody features, and the last layers learn more task-dependant semantic and syntax level features~\cite{10031189, ollerenshaw21_interspeech, ollerenshaw2022}. The research focuses on embeddings at the initial layers of ASR models. Furthermore, we introduce an acoustic embedding-to-waveform synthesis model to synthesise the corresponding audio waveform of the acoustic embedding for better understanding and interpretation (as shown in Stage 2 in Figure~\ref{fig:arc}).

\noindent The main contributions of this paper are as follows:
\begin{enumerate}[itemindent=0.01cm,labelsep=0.01cm]
    \item We propose a method to use single gradient reversal at flexible layers of an ASR model to effectively mitigate speaker information from the representations generated by initial layers of the model without increasing its WER. In the analyses, we observed that speaker identification accuracy was reduced by 22\%  at layer 3 (CE3), 7.3\% at layer 5~(CE5), and 6\% at layer 7~(CE7) compared to the original speech waveform (Table \ref{tab:spk}). Performance of the models trained with these representations was improved by 8.6\% WER on average.    
    The proposed method does not require computationally expensive voice-conversion/speech-synthesis models for anonymization and operates on ASR embeddings. 
    
    \item Our results show that while having improved ASR performance, the speaker adversarial training has anonymized acoustic embeddings with gradient scaling. A detailed analysis of the effects of gradient scaling, domain loss scaling and model layer hierarchies are presented with performance of models and their  convergence properties. Furthermore, the mutual speaker information (depicted in Stage 3  in Figure~\ref{fig:arc}) among the speaker embeddings are analysed and presented.
    \item  Contrary to the previous claims~\cite{Comanducci2021ReconstructingSF}, we show that acoustic embeddings can be re-synthesised to intelligible audio recordings irrespective of certain types of convolution or feed-forward layers in network architectures of the models.
 \end{enumerate}

\newcommand{\modelname}{FleGReSA}
\vspace{-2mm}
\section{Flexible Gradient Reversal Speaker Anonymization (\modelname)} 

The proposed framework with ASR model training (Stage 1), and evaluation phases (Stage 2 \& 3) are shown in Figure \ref{fig:arc}.

\subsection{Stage 1 - Training models and Extracting Embeddings}
\vspace{-0.1cm}
\subsubsection{Training ASR Models}
\label{sec:acoustic}

The ASR model training is shown in Figure \ref{fig:arc}a. We used the conformer model \cite{Gulati2020ConformerCT} as the baseline ASR model consisting of conformer blocks. A conformer block consists of layer normalisation, Feed-forward, Multi-Headed Self-Attention and Convolution modules \cite{Gulati2020ConformerCT}. An x-vector~\cite{8461375} speaker classification model is used with the ASR model for speaker anonymization through speaker adversarial training as described below.

\vspace{-0.1cm}
\subsubsection{Speaker Adversarial Training (SAT)}
\label{sec:adv}
\vspace{-0.1cm}

The SAT aims to learn speaker invariant representations at different layers, and removes speaker specific information from acoustic embeddings \cite{zhou2022enhancing,ganin2016domain}.
We add gradient reversal layer at different hierarchies of the encoder, with relevant gradient scaling, and make the number of speaker invariant layers flexible.
The gradient reversal is a `pseudo function'~\cite{ganin2016domain} $\mathcal{G}(\cdot)$, which defines (a) forward and (b) backward pass with input $x_{k_f}$ by 
\vspace*{-5pt}
\begin{eqnarray}
  \mathrm{(a)} & \mathcal{G}(x_{k_f}) = x_{k_f} \,\,\, \mathrm{and}  \,\,\,  \mathrm{(b)} & \frac{d\mathcal{G}(x_{k_f})}{d x_{k_f}} = -\alpha \cdot \mathbf{I}
   \label{eq:gradient}
   \vspace*{-20pt}
\end{eqnarray}
where $x_{k_f}$ is the output of the $i^{th}$ layer where the gradient reversal $\mathcal{G}(\cdot)$ is applied, $\alpha$ is the gradient scaling factor, and $\mathbf{I}$ is the identity matrix. In the forward-pass (a), it follows the identity transformation, and in the backward-pass  (b),  it is multiplied by $-\alpha$. 
When gradient reversal is added at the $i^{th}$ encoder block, the ASR model is split in: (1) feature extractor $x_{k_f}=G_f(x_k, \theta_f)$ which comprises the  $1^{st}$ to the $i^{th}$ ASR encoder block; (2) speaker invariant encoder $x_{k_m}=G_m(x_{k_f}, \theta_m)$ defined by the remaining layers in the ASR encoder; and (3) ASR decoder $G_y(x_{k_m}, \theta_y)$. The $k^{th}$ input sample to the ASR model is $x_{k}$, and $\theta_f$, $\theta_m$ and $\theta_y$ denote the parameters in the feature extractor, speaker invariant encoder and decoder, respectively.

The discriminative speaker classifier $G_d(x_{k_f},\theta_d)$, which is used to enforce invariant representations, takes input $\mathcal{G}(x_{k_f})$ where $\theta_d$ denotes its parameters. The ASR model loss $L_y$ and the speaker classifier model loss $L_d$ are defined by \cite{10.1145/1143844.1143891, 10.1109/ICASSP.2018.8462105, speechbrain}.
\vspace*{-3pt}\begin{align}
   L_{y}\left ( \theta _{f}, \theta _{m}, \theta _{y}   \right ) \mkern-5mu= &L_{y}\mkern-5mu\left ( G_{y}\mkern-5mu\left (G_{m}\mkern-5mu\left (G_{f}\mkern-5mu\left ( x_{k}, \theta _{f}  \right ), \theta _{m} \right ), \theta _{y}  \right ),y_{k} \right ) \\
    L_{d}\left ( \theta _{f},\theta _{d}   \right ) = &L_{d}\left ( G_{d} \left ( G_{f}\left ( x_{k}, \theta_{f} \right ), \theta_{d} \right ), s_{k}\right )
    \vspace*{-5pt}
\end{align}
where $y_k$ and  $s_k$ are the transcription label and speaker label for the $k^{th}$ sample, respectively. Hence, the final loss is

\vspace*{-16pt}\begin{align}
L(\theta _f, \theta_m, \theta _y, \theta _d) = \frac{1}{K} &\sum _{k=1}^{K} L_y^k(\theta _f, \theta_m, \theta _y) \nonumber \\ &+ \frac{1}{K}\sum _{k=1}^{K}\lambda \cdot L_d^k(\theta _f,\theta _d) %\tag{3} 
\label{eq:adv}
\vspace*{-8pt}
\end{align}
where the total number of samples is $K$ and $\lambda$ is speaker loss regularizer. The gradient of the loss with respect to the input can be written by (dropping arguments of the losses for clarity)

\vspace*{-10pt}
\begin{align}
    \nonumber \frac{\partial L}{\partial x_{k}} \mkern-5mu = \mkern-3mu \frac{\partial x_{k_f}}{\partial x_{k}} \mkern-3mu \cdot \mkern-3mu \frac{\partial L}{\partial x_{k_f}}\mkern-3mu &= \mkern-3mu \frac{\partial x_{k_f}}{\partial x_{k}} \mkern-3mu \cdot \mkern-3mu \left( \mkern-3mu \frac{\partial L_y}{\partial x_{k_f}} \mkern-3mu + \mkern-3mu \frac{d \mathcal{G}(x_{k_f})}{d x_{k_f}} \mkern-3mu \cdot \mkern-3mu \frac{\partial L_d}{\partial \mathcal{G}(x_{k_f})} \mkern-3mu\right) \\ 
    & = \frac{\partial x_{k_f}}{\partial x_{k}} \cdot \left( \frac{\partial L_y}{\partial x_{k_f}} \boldsymbol{-\alpha \frac{\partial L_d}{\partial x_{k_f}}} \right).
    \label{eq:derivative}
    \vspace*{-8pt}
\end{align}
where the term in bold is the gradient injected for speaker adversarial training. The speaker classifier used in the speaker adversarial training is based on x-vector \cite{snyder2018x} model. Unlike the previous works \cite{Srivastava2019PrivacyPreservingAR,Jaiswal2019PrivacyEM}, the speaker adversarial classifier is not a pre-trained model and it is trained jointly with the ASR model (Stage 1). 
After training, the speaker adversarial classifier is removed from the ASR model where the layers are trained to have speaker invariant acoustic representations, and only $\theta_f$, $\theta_m$, and $\theta_y$ are used for decoding.

\subsection{Stage 2 - Training Embedding-to-audio Synthesis and Speaker Recognition Models}

\subsubsection{Neural Embedding to Speech Synthesis}
\label{sec:gan}
Contrary to the previous methods~\cite{meyer2022speaker,meyer2023anonymizing}, where a voice conversion approach is used to convert the audio to a different speakers voice, we directly  anonymize the acoustic embeddings from the ASR model. The speaker privacy in the  anonymized acoustic embedding is evaluated using speaker classifiers \cite{Srivastava2019PrivacyPreservingAR}.

The \emph{discriminative} speaker classifiers may be sensitive to small changes (e.g., perturbation difference) in embedding spaces among different ASR models \cite{szegedy2013intriguing, goodfellow2014explaining}. Moreover, the same utterances may have different embeddings obtained from different ASR models. Therefore, comparing embeddings provided by different ASR models to achieve speaker privacy is not practical. As a result, an extra stage is added to be able to listen to the audio synthesized from acoustic embeddings.

We propose a method to employ acoustic embeddings for audio synthesis and evaluate the anonymization of the generated acoustic embeddings (Stage 2 in Figure \ref{fig:arc}b). The embedding-audio synthesis model is based on HiFi GAN, and a mixture of multi-period and multi-scale sub-discriminators \cite{melgan2019, hifigan2020}. During inference, it takes embeddings from different layers and produces high resolution audio synthesis. If $x_{k_i}$ is the acoustic embedding obtained at the $i$th layer for waveform input $x_k$, then the synthesised output of the generator $\hat{x}_{k_i} = G_{syn}(x_{k_i})$ has the same dimension as $x_k$. According to \cite{hifigan2020}, the training loss for the embedding-audio synthesis training is the summation of generator loss $L_{mel}$ and sub-discriminator loss $L_{FM}$ given by
\vspace*{-8pt}\begin{align}
 &L_{mel}=\frac{1}{K}\sum_{k=1}^K  || \phi (x_k) - \phi(\hat{x}_{k_i}) ||_1,   \\
 &L_{FM} (D)= \frac{1}{K}\sum_{k=1}^K   \sum_{i=1}^{T}\frac{1}{N_i} || D^i(x_k) - D^i(\hat{x}_{k_i}) ||_1 
\end{align}
\normalsize
where $\phi$ is the function used for calculating spectrogram, $T$ is the total number of layers in a discriminator and $N_i$ is the feature dimension of the $i$th layer output denoted by $D^i$. 
 Mel-spectrogram loss ($L_{mel}$) and feature matching loss ($L_{FM}$) calculate $\ell_1$ distances between spectrograms and those between discriminator outputs during training.
The discriminator in the synthesis module comprises $Q$ sub-discriminators $\{D_q\}_{q=1}^Q$ which are used in the final losses:
\vspace*{-5pt}\begin{align}
    L_{G_{syn}} =&\sum_{q=1}^{Q} \Bigg [ \frac{1}{K}\sum_{k=1}^K \left [ (D_q(x_k)-1)^2 + (D_q(\hat{x}_{k_i}))^2 \right ] \nonumber\\ &+ \lambda_{FM}L_{FM} (D_q) \Bigg ] + \lambda_{mel}L_{mel}, \\
    L_D = &\sum_{q=1}^{Q} \frac{1}{K}\sum_{k=1}^K\left [ (D_q(\hat{x}_{k_i})-1)^2 \right ]
\end{align}
\normalsize
where $\lambda_{mel}$ and $\lambda_{FM}$ are loss scaling parameters. 

\subsubsection{Training Speaker Embedding and Identification Models}
\label{sec:spk}

An x-vector model \cite{snyder2018x} pre-trained on Voxceleb1 \cite{nagrani2017voxceleb} and Voxceleb2 \cite{chung2018voxceleb2} is fine-tuned on LibriSpeech data for learning speaker representations. This model shown in Figure~\ref{fig:arc}c is only used to evaluate the synthesized acoustic embeddings for speaker identification performance.

\subsection{Stage 3 - Speaker Anonymization Evaluation}
Speaker anonymization is evaluated on the waveforms synthesized from the acoustic embeddings using the generator described in Section \ref{sec:gan} as depicted in Figure~\ref{fig:arc}d. The acoustic embeddings are obtained from different layers of the ASR model and the waverforms are evaluated with the fine-tuned x-vector model (Section \ref{sec:spk}).

\section{Experimental Setup}

\subsection{Experimental Setup}

\textbf{Data:} The publicly available LibriSpeech~\cite{7178964} corpus has been used for ASR model training (in Figure \ref{fig:arc}a), embedding extraction (in Figure \ref{fig:arc}a) and embedding-audio synthesis (in Figure \ref{fig:arc}b). The train-clean-100 (100 hours) split has been used for training. The \emph{dev-clean}, \emph{test-clean}, and \emph{test-other} splits have been used for validation and testing. Additionally, we have combined train-clean-100 and train-clean-360 into train-clean-460 (460 hours clean speech). This combined set has been used for the training of embedding-to-audio synthesis.

For the speaker adversarial ASR training, the labels for the ASR and speaker classifier models are necessary. The speaker classifier model requires same speakers for training and evaluation. Therefore, before the training, some utterances have been randomly selected and separated from training data for each speaker to create~(\emph{test-adv}).
The speaker classifier shown in Figure \ref{fig:arc}c is fine-tuned with dev-clean, 70\% of the speaker, leaving 30\% for evaluation ({\it dev-clean-te}).

\textbf{Setup:} We performed the experiments in three stages. In the first stage (Figure \ref{fig:arc}a), an ASR model is trained with speaker adversarial loss. In the second stage (Figure \ref{fig:arc}b), acoustic embeddings are extracted from different layers, and then the embedding-audio GAN model is trained to reconstruct the original audio. The hyperparameters for the GAN training are similar to \cite{hifigan2020} \emph{V1}. The synthesis models are trained with the clean 460 hours of LibriSpeech data. In the third stage (Figure \ref{fig:arc}c), the embedding-audio GAN generator is used to synthesize audio from acoustic embeddings to evaluate the speaker anonymity compared to the original audio utterances and baseline. The second and third stages are evaluation stages. The experiments were implemented using Speechbrain \cite{speechbrain}.

\textbf{Baseline:} A conformer \cite{Gulati2020ConformerCT} model with 12 encoder and 4 decoder blocks has been used as the ASR baseline model. The model has $13.3M$ trainable parameters and it is decoded with a language model \emph{shallow fusion} \cite{speechbrain}, beam size 1. The baseline model is used both for training the ASR model and extracting embeddings for audio synthesis. The baseline model embeddings are compared with the {\modelname} embeddings for evaluating their anonymity compared to the original audio samples.

\vspace{-2mm}
\subsection{Evaluation}
\label{sec:eval}

The ASR model is evaluated using Word Error Rate~(WER), and speaker classifier is evaluated using the unweighted accuracy~(WA) metric. ASR performance is evaluated with models where gradient reversal layers are applied at their different layers with different scaling $\alpha$ and $\lambda$ values. The goal is to analyse the impact of gradient reversal, and stabilise ASR training with scaling weights in different layers when gradient reversal is applied. The ASR decoding setup is same as the baseline. The speaker anonymization of the acoustic embeddings obtained from different layers of the ASR model is evaluated using the speaker identification accuracy based on x-vector as mentioned in Section \ref{sec:spk}.

\vspace{-3mm}
\section{Results \& Discussion}

\SetTblrInner{rowsep=0pt}
\SetTblrInner{colsep=2pt}
\begin{table}
\centering
\resizebox{0.9\columnwidth}{!}{
\begin{tblr}{
  cells = {c,m},
  cell{2}{1} = {r=20}{},
  cell{3}{2} = {r=6}{},
  cell{9}{2} = {r=5}{},
  cell{14}{2} = {r=2}{},
  cell{16}{2} = {r=3}{},
  cell{19}{2} = {r=3}{},
  vlines,
  hline{1-2,22} = {-}{},
  hline{3,9,14,16,19} = {2-7}{},
  hline{4-8,10-13,15,17-18,20-21} = {3-7}{},
}
{\footnotesize \textbf{Train} \\ \textbf{Data}} & \textbf{\footnotesize GRL} & \textbf{$\alpha/\lambda$} & {\footnotesize\textbf{test-adv}\\\textbf{\scriptsize WER (\%)}} & {\footnotesize \textbf{dev-clean}\\\textbf{\scriptsize WER (\%)}} & {\footnotesize \textbf{test-clean}\\\textbf{\scriptsize WER (\%)}} & {\footnotesize \textbf{test-other}\\\textbf{\scriptsize WER (\%)}}\\
{train\\clean\\100} & \textbf{-} & \textbf{-} & \textbf{-} & \textbf{6.14} & \textbf{6.18} & \textbf{16.28}\\
 & CE3 & 0.01/0.5 & 4.55 & 5.38 & 5.58 & 16.27\\
 &  & 0.1/0.3 & 3.23 & 5.48 & 5.72 & 15.94\\
 &  & 1.5/0.3 & 3.48 & 6.19 & 6.77 & 18.09\\
 &  & 1.0/0.1 & 3.51 & 6.06 & 6.47 & 18.27\\
 &  & 0.5/0.5 & 3.08 & 5.76 & 6.23 & 17.63\\
 &  & 1.0/0.05 & 3.85 & 5.68 & 5.92 & 17.57\\
 & CE5 & 0.5/0.5 & 2.81 & 6.27 & 6.74 & 18.31\\
 &  & 0.1/0.3 & 3.73 & 5.34 & 5.75 & 15.96\\
 &  & 0.01/0.3 & 3.69 & \textbf{5.26} & \textbf{5.47} & 16.04\\
 &  & 0.5/0.05 & 2.78 & 5.54 & 5.80 & 16.85\\
 &  & 1.0/0.1 & 3.36 & 5.89 & 6.27 & 17.36\\
 & CE7 & 1.0/0.05 & 4.29 & 5.78 & 6.11 & 17.55\\
 &  & 0.05/0.3 & 3.69 & 5.41 & 5.64 & \textbf{15.92}\\
 & CE10 & 1.0/0.3 & 3.59 & 5.95 & 6.31 & 17.46\\
 &  & 0.1/0.3 & 4.02 & 5.47 & 5.88 & 17.09\\
 &  & 0.5/0.5 & 4.12 & 6.12 & 6.71 & 17.93\\
 & CD4 & 0.5/0.3 & 3.27 & 5.60 & 5.94 & 17.30\\
 &  & 0.5/0.5 & 4.21 & 5.63 & 5.87 & 17.32\\
 &  & 1.0/0.1 & 4.12 & 5.42 & 6.10 & 17.02
\end{tblr}
}
\caption{An analysis of the ASR performance (WER) applying gradient reversal at different layers of the ASR model with different $\alpha$ and $\lambda$, where \emph{GRL} denotes the gradient reversal layer.}
\vspace*{-30pt}
\label{tab:adv_asr}
\end{table}

The ASR performance of the speaker adversarial ASR is shown in Table \ref{tab:adv_asr} where: CE denotes \emph{conformer encoder}; CD denotes \emph{conformer decoder}; the number following CE or CD is the embedding layer number; $\alpha$ and $\lambda$ are scaling factors used in Eq. \eqref{eq:gradient} and \eqref{eq:adv}. Instead of applying the adversarial layer only at the end of the encoder~ \cite{Srivastava2019PrivacyPreservingAR}, we propose flexible speaker adversarial at various hierarchies of the encoder/decoder model and found ASR performance improvements. The \emph{test-adv} WER shows the ASR performance on utterances which have common speakers with the training data but not common utterances. 
The other test sets are standard \emph{dev-clean}, \emph{test-clean} and \emph{test-other}. The overall results given in Table \ref{tab:adv_asr} show that the ASR performance obtained from speaker adversarial training improves across the test scenarios compared to the baseline (first row). We observe that adding GRL in the lower layers does not decrease the ASR performance. The weight of the gradient reversal layer is crucial in the initial convergence and overall performance of the ASR model  \ref{tab:adv_asr}. The results show that that high values for $\alpha$ and $\lambda$ prevent the ASR model from converging. Furthermore, the weight of the gradient reversal layer $\alpha$ is also dependant on the layer of the ASR model where the gradient reversal layer is injected, as the linguistic and speaker information are highly entangled at the initial layers of the encoder of the ASR model~\cite{10031189, ollerenshaw21_interspeech}. The $\alpha$ and $\lambda$ weights need to be smaller to make the ASR model stable as lower negative speaker gradients distort the sequential linguistic entanglement in the acoustic embeddings, and it loses the linguistic boundary information. As a result, the ASR model mostly predicts blanks and misaligned word sequences.

Next, we analyze how the layers can become speaker invariant after the intersection of gradient reversal layers. In Table \ref{tab:spk}, the higher the speaker accuracy, the less anonymous the speaker representations are. The results show with adversarial training, the ASR model embeddings are more speaker redundant. The \emph{adv\_CE3D} model shows when the gradient reversal is at layer 3 (CE3) and the embedding is extracted from layer 5 (CE5), the acoustic embeddings are more anonymous compared to the acoustic embeddings extracted from layer 3. This suggests that we can control the trade-off between embedding speaker quality and downstream task performance  by flexible adversarial training. Thereby, we achieve speaker anonymity in acoustic embeddings without expensive efforts like  voice morphing or conversion \cite{Qian2018HidebehindEV, Shamsabadi2022DifferentiallyPS, meyer2022speaker}. Next, we compare the audio waveform reconstructed using the \emph{baseline} model to the original audio waveform. We observe that plenty of speaker information remains in the acoustic embeddings at the convolution and fully-connected layers obtained from the  \emph{baseline}.

The speaker anonymization of the embeddings is further assessed computing the mutual information (MI) of random variables of the embeddings. For this purpose, we compute the MI using embeddings $\hat{x}_{k_i}^b$ obtained at the $i^{th}$ layer of the baseline model and the embeddings $\hat{x}_{k_i}^a$  obtained at the $i^{th}$ layer of the anonymized model. The MI is computed  between the original waveform $x_k$ and the synthesized audio $\hat{x}_{k_i}$ using
\vspace*{-5pt}\begin{align}
    \label{eq:mi}
    \mathcal{I}(x_k,\hat{x}_{k_i})\mkern-5mu=\sum_{x_k, \hat{x}_{k_i}}  \mkern-5mu p(x_k,\hat{x}_{k_i})\log\frac{p(x_k,\hat{x}_{k_i})}{p(x_k)p(\hat{x}_{k_i})}.\vspace*{-18pt}
\end{align}
\vspace*{-5pt}

\begin{figure}[] 
\vspace*{-0pt}
\subfloat[]{\includegraphics[width=0.5\columnwidth,keepaspectratio]{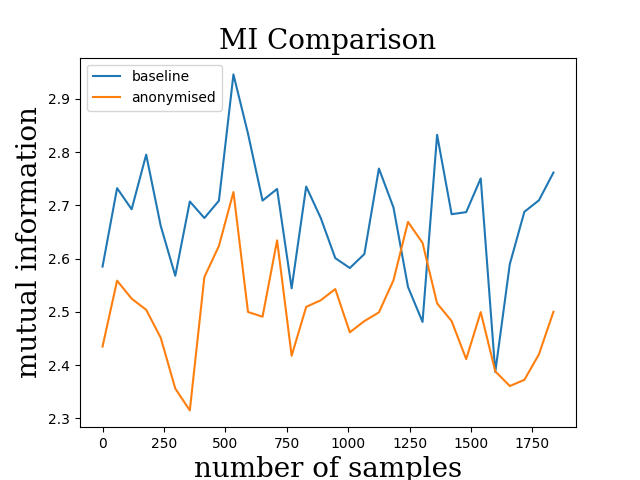}\label{fig:micomp}}
% \hspace{\fill}
\subfloat[]{\includegraphics[width=0.465\columnwidth,keepaspectratio]{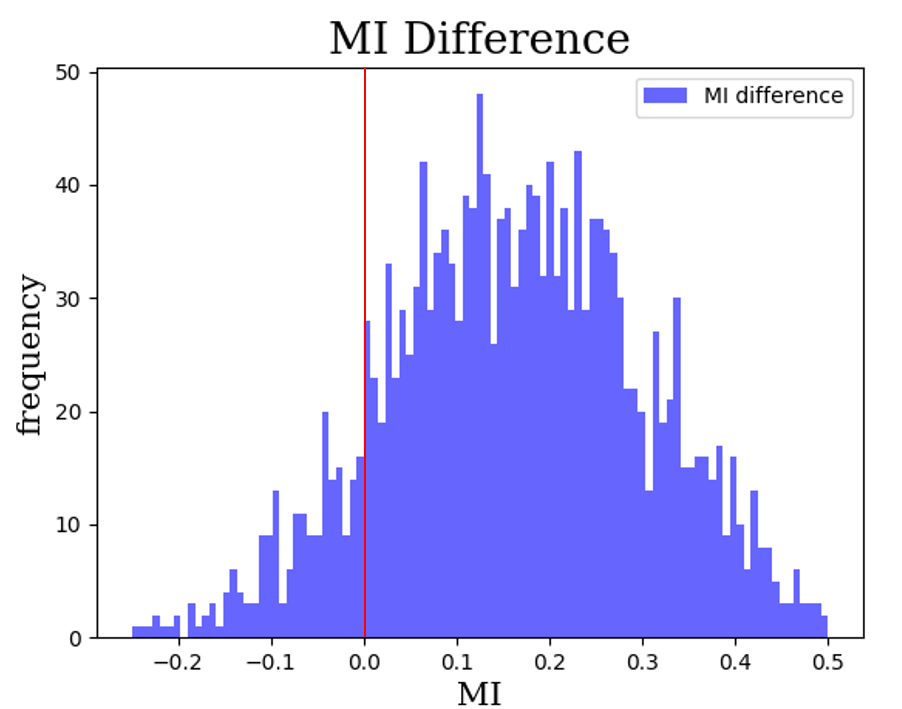}\label{fig:mihist}}
\vspace*{-8pt}\caption{(a) Comparison of MI computed using speaker embeddings obtained from baseline and FleGReSA. (b) Frequency of difference of the MI. }\label{fig:mi}
\end{figure}

The frequency of the MI difference $\mathcal{I}(x_k,\hat{x}_{k_i}^b) - \mathcal{I}(x_k,\hat{x}_{k_i}^a)$ is plotted as a histogram to analyze the information loss  among samples in Figure \ref{fig:mi}. In Figure \ref{fig:micomp}, the blue line denotes the speaker MIs computed with $\mathcal{I}(x_k,\hat{x}_{k_i}^b)$ for dev-clean where $\hat{x}_{k_i}^b$ is generated with the baseline synthesised model (i.e. Baseline in Table \ref{tab:spk}). The orange line in Figure \ref{fig:micomp} denotes the speaker MIs calculated as $\mathcal{I}(x_k,\hat{x}_{k_i}^a)$ where $\hat{x}_{k_i}^a$ is generated with the anonymized model (i.e. adv\_CE3D\_v1 in Table \ref{tab:spk}). The difference between these two curves is displayed in Figure \ref{fig:mihist} as a histogram. These  results  show  evidence  of  the  speaker information  reduction  using  the  anonymized  model  where  a substantial proportion of the utterances is reduced after speaker anonymization. These results corroborate the findings observed in Table \ref{tab:spk}.

\begin{table}
\centering
\resizebox{0.99\columnwidth}{!}{
\begin{tblr}{
  cells = {c},
  cell{2}{1} = {r=17}{},
  cell{3}{2} = {r=3}{},
  cell{3}{3} = {r=3}{},
  cell{3}{4} = {r=3}{},
  cell{3}{5} = {r=3}{},
  cell{6}{2} = {r=3}{},
  cell{6}{3} = {r=3}{},
  cell{6}{4} = {r=3}{},
  cell{6}{5} = {r=3}{},
  cell{9}{2} = {r=3}{},
  cell{9}{3} = {r=3}{},
  cell{9}{4} = {r=3}{},
  cell{9}{5} = {r=3}{},
  cell{12}{2} = {r=2}{},
  cell{12}{3} = {r=2}{},
  cell{12}{4} = {r=2}{},
  cell{12}{5} = {r=2}{},
  cell{14}{2} = {r=3}{},
  cell{14}{3} = {r=3}{},
  cell{14}{4} = {r=3}{},
  cell{14}{5} = {r=3}{},
  cell{17}{2} = {r=2}{},
  cell{17}{3} = {r=2}{},
  cell{17}{4} = {r=2}{},
  cell{17}{5} = {r=2}{},
  vlines,
  hline{1-2,19} = {-}{},
  hline{3,6,9,12,14,17} = {2-7}{},
  hline{4-5,7-8,10-11,13,15-16,18} = {6-7}{},
}
{\textbf{\textbf{Train}}\\\textbf{\textbf{ Data}}} & \textbf{Model} & \textbf{ GRL} & {\textbf{dev-clean}\\\textbf{ WER(\%)}} & \textbf{$\alpha / \lambda$} & \textbf{AE} & {\textbf{dev-clean-te}\\\textbf{ SPK-ACC}}\\
{train\\clean\\100} & Original audio & - & - & - & - & 96.9\\
 & Baseline & - & 6.14 & - & CE3 & 86.6\\
 &  &  &  &  & CE5 & 22.1\\
 &  &  &  &  & CE7 & 6.4\\
 & ~ ~adv\_CE1 & CE1 & 5.58 & 0.5/0.05 & CE3 & 73.9\\
 &  &  &  &  & CE4 & 41.7\\
 &  &  &  &  & CE5 & 42.9\\
 & adv\_CE3D\_v1 & CE3 & 5.76 & 0.5/0.5 & CE3 & \textbf{64.6}\\
 &  &  &  &  & CE4 & \textbf{33.8}\\
 &  &  &  &  & CE5 & \textbf{14.8}\\
 & adv\_CE3D\_v2~ & CE3~ & 6.06 & 1.0 / 0.1 & CE3 & 71.1\\
 &  &  &  &  & CE5 & 18.0\\
 & adv\_CE5D & CE5 & 5.54 & 0.5/0.05 & CE5 & 18.2\\
 &  &  &  &  & CE6 & 0.8\\
 &  &  &  &  & CE7 & \textbf{0.4}\\
 & adv\_CE10D & CE10 & 5.48 & 0.5/ 0.3 & CE3~ & 68.8\\
 &  &  &  &  & CE5 & 70.2
\end{tblr}
}
    \caption{Speaker accuracy on the re-synthesised waveforms from acoustic embeddings at different layers, where \emph{AE} denotes the acoustic embedding extraction point and \emph{SPK-Acc} denotes the unweighted speaker accuracy (\%).}\vspace*{-15pt}
\label{tab:spk}
% \vspace{-5mm}
\end{table}

\section{Conclusion}
In this paper, a flexible gradient reversal speaker anonymization (\modelname) and evaluation framework is presented. One of the main benefits of the proposed framework is performing anonymization as an integral part of the ASR model. Once we train the  ASR model with the domain adversarial speaker classifier, the latter is discarded. The ASR model is solely employed to provide the anonymous acoustic embeddings. We showed that the training is flexible depending upon the acoustic embedding extraction layer and desired downstream task. The results show that the ASR model is stable and performs better with the adversarial training, while providing significant speaker anonymization on the acoustic embeddings. 
Experimental results obtained using the LibriSpeech indicate that in the best case the proposed approach achieves a remarkable reduction in speaker recognition accuracy by an absolute 22\%. Furthermore, the best ASR performance among the models improves the relative WER of the ASR model by 14\%. Furthermore, we have presented an embedding to audio  high-quality waveform synthesis model not only comparing speaker information but subjectively listening to the synthesized audio of layer-wise embeddings.

\clearpage
\bibliographystyle{IEEEtran}
\bibliography{mybib.bib}

\end{document}